# Nanomaterials decoration on commercial cotton bandages for pain and infection management

Rohit Parkale[1], Priyanka Pulugu[1], Prasoon Kumar[1,2]*

[1]Department of Medical Devices, National Institute of Pharmaceutical Education and Research (NIPER)-Ahmedabad, Palaj, Gandhinagar-382355, Gujarat, India.

[2]Department of Biotechnology and Medical Engineering, National Institute of Technology, Rourkela 769008, Odisha

*corresponding author: prasoonkumar1985@gmail.com

**Abstract:**

Cotton gauze bandages (CGB) are one of the most widely used wound dressing materials. These bandages are placed over a wound site to keep it clean and facilitate the healing process. However, it is used along with antibacterial agents (ointment) to prevent post-dressing infections. Further, other medications have to be orally administered for managing pain and wound healing. Therefore, in the current work, nanomaterials having drug releasing and antibacterial properties is coated onto the cotton gauze bandage to minimize the local pain and post-dressing infections at a wound site. We fabricated diclofenac sodium (an anti-inflammatory drug) loaded biodegradable, polycaprolactone (PCL) nanofibers mat through electrospinning and deposited it over the surface of a CGB that was initially coated by chitosan and decorated with ZnO nanoparticles. Chitosan coating over the CGB provides the antibacterial properties while the drug loaded nanofibers mat releases diclofenac sodium under a simulated drug release condition. In addition, the decorated ZnO nanoparticles have ultra violet radiation blocking properties. The modified bandage was characterised by SEM, EDAX, DSC, UV spectroscopy, and FTIR. The antibacterial property of the bandage was evaluated through zone of inhibition test during the anti-bacterial study. Thereafter, drug release and degradation studies revealed the modified bandage could provide sustained drug release for 15 days, perhaps through surface erosion and later bulk degradation mechanism. Thus, the nanomaterials decorated CGB can be a better alternative to native CGB for wound dressing applications.

***Keywords*:** Nanofiber, Drug delivery, Wound dressing, Antibacterial, UV blocking, ZnO nanoparticle.

## 1.0 Introduction

Skin wound dressing materials form an important segment of wound care industry. The market is expected to grow at the rate of 20% from 2016 to 2020 [1]. Out of these wound dressing materials, commercial cotton gauze happens to be the most popular of the surgical wound dressing materials due to readily available, economical, highly absorbent and nature [2]. However, CCG application leads to local evaporative cooling at a wet wound tissue, thereby lowering the temperature of the wound. At low temperature, there is an impaired would healing due to increased affinity of haemoglobin for oxygen created by hypoxia, compromised leukocyte and phagocyte activity increases the chances of infection and reflex vasoconstriction. In addition, on removal of CCG dressing from the wound site, patient experiences discomfort and pain due to non-selective mechanical debridement of adjacent healthy tissue. This not only affects the wound care method but also makes the wound vulnerable to cross-contamination by dispersion of bacteria. Thus, the cost and average nursing time gets significantly increased for patient as CCG dressing requires dressing changes thrice a day [3]. Hence, there is a need to develop an alternate wound dressing material or modify the conventional CCG to overcome its limitations.

Several researchers have demonstrated that the surface modification of CGB by using natural polymers like chitosan or heavy metals like ZnO, Ag can reduce infection rate and hence, accelerate wound healing process [4, 5] [6]. The use of chitosan is preferred to heavy metal in homeostasis and antibacterial activity due to local toxicity issue associated with sometime heavy dose of heavy metals [7]. In addition to the wound healing support, efforts have been directed towards localized pain management at these wound site through sustain delivery of analgesic drugs like Diclofenac sodium from biodegradable polymeric(Polycaprolactone) nanofibers patches[5]. However, standalone use of these nanofiber patches is challenging, as they require substrate materials for imparting them with sufficient mechanical strength during handling and usage. Nevertheless, nanofiber patches has been observed to be an effective in chronic wounds like diabetic foot ulcers [8]. The discomfort and pain associated to mechanical debridement is prevented in nanofiber patches as they get absorbed at the wound site. However, such wound patches become an expensive alternative to CCG bandages. Therefore, a cost effective modification of CCG using chitosan nanoparticles and drug loaded nanofibers can provide a dual advantage of minimizing infection to enhance the healing time of wound and manage localized pain at wound site[9, 10]

In the present study, we modified the conventional CGB using nanomaterials for antibacterial protection and sustained transdermal drug delivery action. The CGB fabrics is coated by 2% chitosan nanoparticles, ZnO nanoparticles and decorated with diclofenac sodium loaded Polycaprolactone (PCL) nanofibers. This modified CCG bandage has antibacterial, ultra violet radiation blocking properties and localised transdermal drug delivering properties. This type of smart modified wound dressing materials can prevent the infections, bleeding and pain over the wound site.[11].

## 2.0 Materials and Methods
### 2.1 Materials
Span 80 (Sigma Aldrich), Glycerol (HI Media), Hydrogen peroxide (Fischer Scientific), Zinc Oxide (Sigma Aldrich), Chitosan (Sigma Aldrich), Diclofenac Sodium (Swapnaroop Drugs and Pharmaceuticals, India), Polycaprolactone (80 kD - Sigma Aldrich), Glacial Acetic acid (Fisher Scientific). Sweat Simulated Fluid Composition (g/L) (pH 4.5) - Sodium Chloride: 5.49

(Fisher Scientific), Calcium Chloride: 3.32 (Fisher Scientific), Magnesium Sulphate:0.24 (Sigma- Aldrich), Potassium Dihydrogen Phosphate: 1.36 (EMPLURA).

## 2.2. Modification of marketed cotton gauze-
### 2.2.1. Chitosan coating

Commercial cotton gauze bandage obtained from the local market was thoroughly cleansed to remove pollutants. The fabrics were boiled in a mixture of 4.0 g/L of sodium hydroxide, 1 g/L of glycerol and 2.0 g/L of Sorbitan monooleate (span80) for a duration of 50 minutes to remove dust, fatty substances, pigments and other natural impurities. Thereafter, the fabrics were bleached again in the above mixture of ingredients before treating with hydrogen peroxide (35%) at $80^0$C for 30 minutes to remove the natural pigmentation from the fabrics until a permanent whiteness on the surface was obtained. The chitosan coating over the fabrics was achieved by dipping these cleansed cotton fabric gauzes into 2% chitosan solution in an aqueous acetic acid for 8 hours. Thereafter, these cotton fabric samples were removed from the solution and the excess of the absorbed aqueous acetic acid was drained out via compression. Finally, the cotton fabric gauzes were dried at $60^0$C for 20 minutes. The coating of chitosan was confirmed by Fourier Transform Infrared Spectroscopy (FT- IR) (Alpha Opus 7.5, Bruker) and Field Emission scanning electron microscopy (FE-SEM)(Zeiss Laboratory) [12].

### 2.2.2. ZnO nanopowder over the fibril of bandage

The 10ml of 0.5mg/ml solution of ZnO nano-sized powder was added to the pre-weighed bandage. Thereafter, the fabrics was submerged into the ZnO solution and agitated on an orbital shaker (Scigenics Biotech, Chennai) at 80rev/min for 24hr. Finally, the loading of ZnO Nano nanoparticle into the fabrics was confirmed by (FE-SEM) and EDS (Energy Dispersive X-Ray Spectroscopy) report.

### 2.2.3. Electrospinning of drug-loaded nanofibers-

The solution of 10% polycaprolactone (PCL) wt./vol. was prepared in chloroform. [13]. The solution was homogenized by stirring it over a magnetic stirrer (IKA® RH digital) for 6 hours. Then 2% Diclofenac Sodium solution prepared with solvent methanol was slowly added into PCL solution while stirring the solution at 500rpm. The 3:1 ratio of Chloroform: Methanolic drug solution was maintained and was further homogenized for 2 hours [14]. This solution was used to spun nanofibers by an electrospinning technique (ESPIN NANO, Physics equipment's). The parameters were optimized to fabricate drug loaded nanofiber mesh. The parameters include electrode distance of 25cm, voltage - 10KV, flow rate - 0.5ml/hr. and needle gauze of 21G. This drug loaded nanofiber mesh was deposited on the previously fabricated chitosan coated and ZnO decorated cotton gauze pads. The diameter of the nanofibers and their deposition over the CGB was characterized by analyzing through FE-SEM (Zeiss Laboratory) images obtained at different magnification. The nanofibers were observed at the accelerating voltage of 2-3KV. The characterization of drug loaded nanofibers was further performed via FT-IR spectroscopy and Differential Scanning Calorimetry analysis (DSC 214, Polyma, NETZSCH).

### 2.2.4. Drug release and Degradation studies-

*In-vitro* drug release study was performed by using Sweat Simulated Fluid (SSF). The setup was designed as shown in the **Figure S1 A** to study the drug release from the nanofiber mat. The setup includes an elevated platform onto which a piece of cotton gauze having drug loaded nanofiber was placed. The elevated platform is placed in a 30ml pool of SSF in a petri plate such that the two ends of cotton gauze piece are submerged in SSF. After the cotton gauze absorbed SSF till saturation, an aliquot of 1ml were taken out from the SSF pool at regular intervals (1,2, 3…,12,15 days), for estimation of released drug using Ultra Violet-Visible spectroscopy (UV-1800, Shimadzu, Japan). The samples were scanned in the range of 200-400nm wavelength during UV–Vis spectroscopy and absorbance and respective concentration of drug was noted[15, 16]. Thereafter, again a fresh 1ml of SSF media was added to the petri plate to equilibrate the volume of SSF media. A characteristic peak at 276nm was preferred to analyze the release pattern of the drug as shown in the **Figure 2A**. The final amount of drug release were calculated from the standard curve (Abs vs Conc).

$$\% \text{ drug release was calculated by } = \frac{\text{Total amount of drug}}{\text{Total amount of drug present initially}} \times 100$$

The % of drug release was calculated by assuming the drug is uniformly distributed over the nanofiber mesh.

4.19mg of samples was placed in SSF (Sweat Simulated Fluid) media at 37°C in an incubator (Orbitek Chennai). The SSF was changed after each day. The samples was collected after every 24hr. before weighing, the sample was washed with 0.01M NaCl salt solutions and dehydrated. The mass of dehydrated samples were evaluated and the percentage of weight loss were determined by the given formula [11, 17].

$$\% \text{ Degradation of nanofibrous film} = \frac{\text{Initial weight} - \text{Final Weight}}{\text{Initial weight}} \times 100$$

### 2.2.5. Fourier Transform Infrared spectroscopy

FTIR was performed to confirm the chitosan coating over the fabric and the drug loading into nanofiber mat. The sample was prepared by mixing and triturating the potassium bromide (KBr) and minute quantity (1/10$^{th}$) of pure drug. The above process was repeated to prepare sample of drug-loaded sheet and chitosan-coated fabric for FTIR analysis. IR spectra were scanned over a wavenumber range of 4000-400 cm$^{-1}$ [18].

### 2.2.6. Differential Scanning Calorimetry

The thermal characteristic of nanofiber mesh was determined by differential scanning calorimetric (DSC) thermogram analysis. All samples were weighed directly in an aluminum pan (2.5 mg, nanofiber, n=3) and then scanned in the temperature ranging from 10-300$^0$C at a heating rate of 10 $^0$C/min. The nitrogen gas was used at a flow rate of 70ml/min. Further DSC thermogram of PCL and drug loaded PCL fiber extracted in the temperature range of 20-280 $^0$C [19].

### 2.2.7. UV Blocking activity of fabric

The cotton gauze fabric soaked with ZnO nanoparticles solution was tested for UV protection activity. It was evaluated by UV absorption and transmittance through ZnO nanoparticle solution. The transmittance decreases from 265-400nm, which shows that fabric loaded with ZnO solution has UV A and UV B protection (tanning rays) activity [20].

### 2.2.8. Antibacterial test

Antibacterial activity of chitosan (Low Molecular Weight and Medium molecular weight) coated was evaluated by disc diffusion assay. The uncoated fabric was taken as a control for this study. *S.Aureus* and, *E. Coli* are used as standard bacterium for the study of antimicrobial potential of chitosan decorated fabric. Mueller Hinton (MH) Agar medium was prepared which is generally used to identify the susceptibility of non-fastidious bacteria. The inoculum was prepared by inoculating a plate with standard ATCC (American Type Culture Collection) culture and allow it to incubate at 37°C for 24 hr. in an incubator (Thermo scientific). Thereafter, the bacterial colonies were added to a saline suspension and compared the resulting suspension with McFarland standard (turbidity standard). Streaking the swab (Sterile swab dip into inoculation tube) on MH agar coated petri plate; to ensure the uniformity of the inoculum the plates were rotated at 60 °C. Finally, cotton and chitosan coated fabrics samples were placed gently on the agar surface with a sterile forceps Once all disks are in place, replace the lid, invert the plates, and place them in a 37°C incubator for 24hrs[21, 22].

## 3. Results and discussion
### 3.1. Characterization of chitosan coating on CGB

FE-SEM revealed the topographical changes on the cotton fibril before **(Figure 1A)** and after coating with chitosan solution. After coating of chitosan, the fibrils are well separated and can be observed in the **Figure 1 B**. To evaluate the type of bonding between the cotton fabric and chitosan, FT-IR analysis of native cotton fabric, chitosan powder, and chitosan coated cotton fabric was performed **(Figure 2B)** [12]. The FTIR spectra of plane cotton fabric (A) shows a broad band of stretching vibration of OH group from 3200-3438$cm^{-1}$. The sharp peak is observed around 2909 $cm^{-1}$ due to stretching vibration of $CH_2$ group. The remaining peak shows at 1635$cm^{-1}$, 1425$cm^{-1}$ and 1163$cm^{-1}$ belong to O-H bending, $CH_2$ bending and C-O-C stretching, respectively. In the spectra of chitosan (B) shows a broad peak at 3408 $cm^{-1}$ shows the stretching vibration of the OH and NH group. At 1643$cm^{-1}$ shows NH bending due to C=O group extinction. The peak shown at 1151$cm^{-1}$ conform the C-O-C bridge asymmetric stretching. The spectra of chitosan coated cotton fabric (C) shows a broad peak at 3451$cm^{-1}$ belongs to O-H and N-H group stretching. The sharp peak at 1631$cm^{-1}$ is characteristic peak of C=O of amide. The peak from the range of 1600-1500$cm^{-1}$ is belong to N-H bending and the peak shown at 1417$cm^{-1}$ shows the C=N stretching as shown in **Figure 2B** [23]. A prominent peak at 1631 $cm^{-1}$ for cotton fabric (chitosan treated) reveals that the formation of C=N double bond (Schiff base) between aldehydic carbonyl group of cotton and amino group of chitosan.

### 3.2. Characterization of ZnO nanoparticle coating on CGB

The ZnO nanoparticles loaded onto the fibrils of cotton bandage is observed in FE-SEM images. The uniform decoration of ZnO nanoparticles onto the cotton fabric can be observed as shown in the **Figure 1C**. The EDAX spectroscopy analysis suggests 19.95% of ZnO by weight on the fibril of cotton gauze (**Figure S1 B**). The elemental analysis shows the peak of Zn and O have a concentration of 20% and 19% by weight. This confirms the presence of ZnO over the fibrils of cotton bandage. It was found that the average size of ZnO particles was found to be 211nm as shown in the **Figure SI C** [20]. This size range significantly decrease the transmittance of UV rays and provide UV blocking property to the CGB[24]

### 3.3. Characterization of drug loaded nanofibers

FE-SEM images revealed that the diclofenac sodium loaded nanofiber membranes were having random morphology, beadless and continuous fibers. FE-SEM confirms the formation and collection of nanofibers (**Figure 1C**), drug loaded nanofibers (**Figure 1D**) and drug loaded nanofibers over the cotton fabric (**Figure 1E**). EDAX report of drug loaded nanofiber reveal the elemental composition of drug loaded nanofiber EDAX report of drug loaded nanofiber as shown in **Figure S1 D**. The average diameter of PCL nanofiber was found to be the 110.08±6.08 nm size while the diameters of drug-loaded nanofibers were 216.36±4.51 nm higher than the PCL nanofiber (graph showing the comparison of fiber diameters as discussed Figure 1) To confirm the drug loading and understand the interaction between PCL and diclofenac sodium, FTIR analysis was performed for plain PCL nanofiber sample, pure diclofenac sodium and drug-loaded PCL nanofiber (**Figure 2C**). The spectra of the plain PCL shows the characteristic vibration of the $CH_2$ and $C=O$ group at 2980 $cm^{-1}$ and 1722 $cm^{-1}$, respectively PCL spectra also show $CH_2$ bending vibration at 1300$cm^{-1}$ and 1181$cm^{-1}$ shows the characteristic stretching of the ester COO group. 729$cm^{-1}$ shows rocking vibration of $CH_2$ group. In spectra B of diclofenac, it shows an intense defined peak around 1600 $cm^{-1}$($C=C$ peak) and 3280$cm^{-1}$ shows N-H stretching. Spectra C completely revealed the loading and interaction between PCL and diclofenac sodium. The N-H intense peak shift to high wavelength and characteristic peak at 1738 $cm^{-1}$ shows the loading of the drug into a nanofiber as shown in **Figure 2C**[25]. DSC thermograms of drug, polymer, polymer, and drug are shown in the **Figure 2D** Pure drug (DS) and polymer (PCL) show a strong endothermic peak at 284.6$^0$C and 59.1$^0$C respectively. The drug-loaded PCL didn't show an endothermic peak at 284. This reveals that Diclofenac sodium has the highest crystallinity however, once loaded into PCL nanofiber during electrospinning, it may have undergoes amorphization [26]. Amorphization enhances solubility and useful in the desired release of the drug [27].

### 3.4. Functional evaluation of nanomaterial decorated CGB

The drug was released from the PCL nanofiber mesh by diffusion-based manner. Initially, the erosion of polymer layer starts due to acidic SSF fluid and subsequently, drug was released. Initially, a burst release of drug was observed till 4-5 days of experimental setup and thereafter, drug release pattern showed sustained behavior. Eventually, the drug release started decreasing after 9 days as shown in the **Figure 3A**. The drug added into the solution was around 100mg which is a safe dose for topical pain relief activity [28] After 13 days of drug release experiment, the morphology of nanofibrous fabric was observed through FE-SEM as shown in the **Figure 1G**. It was found that the nanofibers degraded to release the drug. The decrease in weight of drug loaded nanofibrous scaffold in presence of sweat simulated fluid is shown in **Figure 3B**. An accelerated decrease in the weight of the nanofiber mesh was observed during later stages of drug release. The degradation ranges from surface erosion of nanofibers to bulk degradation. Following incubation zone has been measured with HI Antibiotic Zone scale (Hi Media). The posterior part of the agar plate was observed with bare eye. Reflected light was used to illuminate the non-reflecting surface with. **The zone of bacterial inhibition was found to be 27 mm and 32mm for gram +Ve and gram –Ve bacteria**. Zone size put down on the recording sheet as shown in the **Figure 3C**[29]. The solution used for the soaking of fabric was tested for UV protection activity by checking the transmittance and absorbance of the ZnO nano powder solution. The transmittance decreases from 265-400nm, which shows that fabric loaded with this solution has UV A and UV B protection (Tanning rays) activity as shown in the **Figure 3D**[30]

### 4. Conclusion

In the current manuscript, we report a smart cotton fabric as a potential next generation dressing material. Here, we coated the CGB with chitosan and decorated the fibrils of cotton fabric by ZnO nanoparticles. The same CGB was also decorated by depositing Diclofenac sodium drug loaded polycaprolactone biodegradable nanofibers. The coated chitosan shows a good antibacterial property and loaded ZnO shows excellent UV screening properties. The sustain drug delivery of the modified CGB can potentially provide the pain relief action over the affected area.

**Credit authorship contribution statement**
Rohit Parkale: Data generation, Data curation and Investigation, Methodology, Visualization Writing of - original draft.
Priyanka Pulugu: Data Analysis, Optimization and inputs in drafting.
Prasoon Kumar: Conceptualization, Investigation, Project administration, Supervision, improvement in the manuscript in terms of conceptualization, Writing - review & editing.

**Declaration of Competing Interest**

The authors declare that they have no known competing financial interests or personal relationships that could have appear to influence the work reported in this paper.

**Acknowledgement**

The authors are highly acknowledged to National Institute of Pharmaceutical Education and Research Ahmedabad for their financial support. We are highly indebted to Dr. Sharda Deore and Dr. Prashant Thakare from Sant Gadage Baba University (SGBAU) Amravati for their cooperation in antibacterial studies.

# Graphical abstract

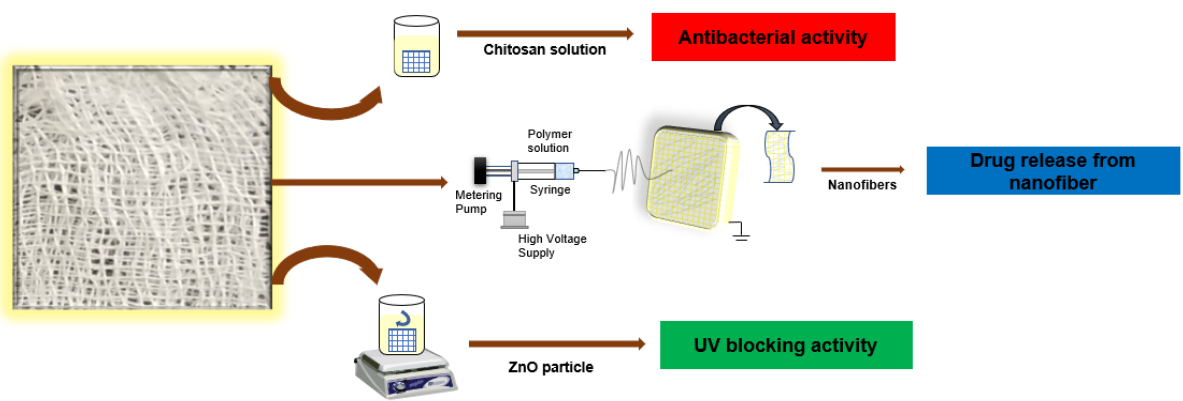

# Figures

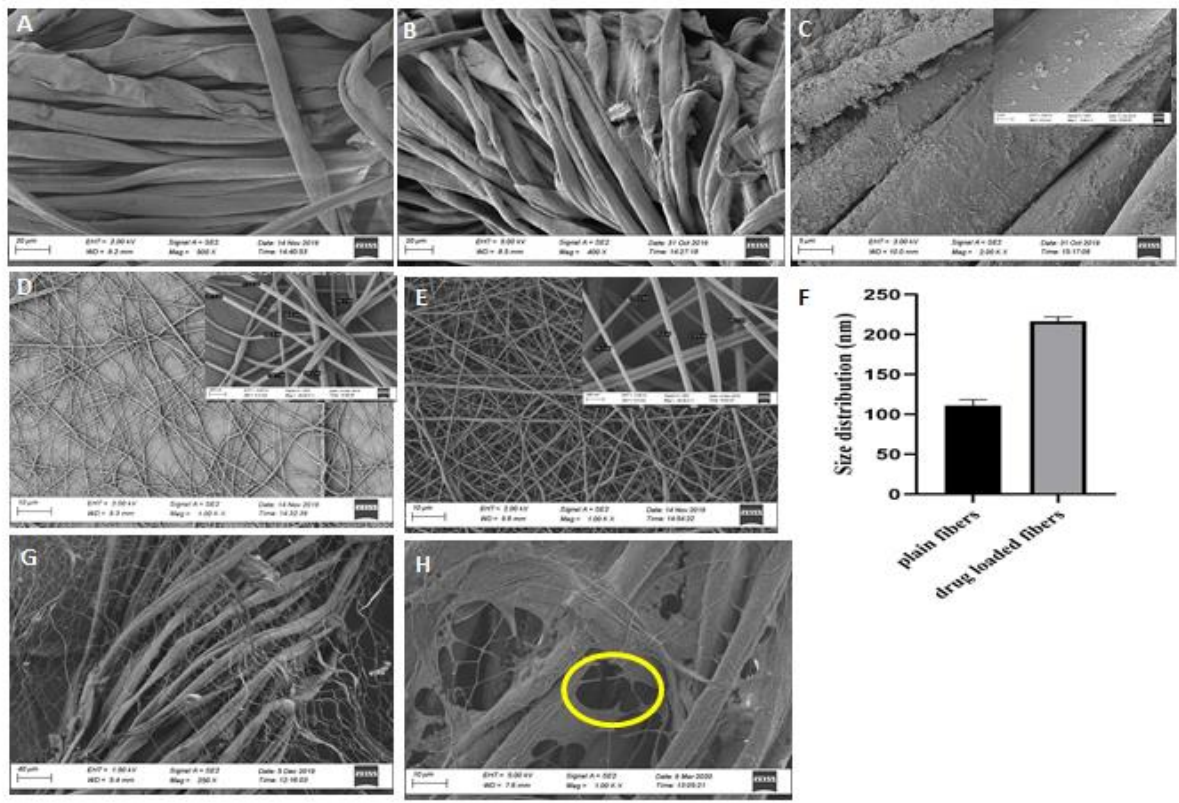

Figure 1 FE-SEM image of A) Plane cotton fabric B) Chitosan coated cotton fabric C) ZnO loaded cotton fiber D) Morphology of PCL nanofiber E) Drug loaded PCL nanofiber. F) Graph showing the comparison of diameter of nanofibers with and without drug loading. FE-SEM image of G) Deposition of PCL nanofiber over the cotton fabric H) Degradation of Drug loaded PCL nanofiber

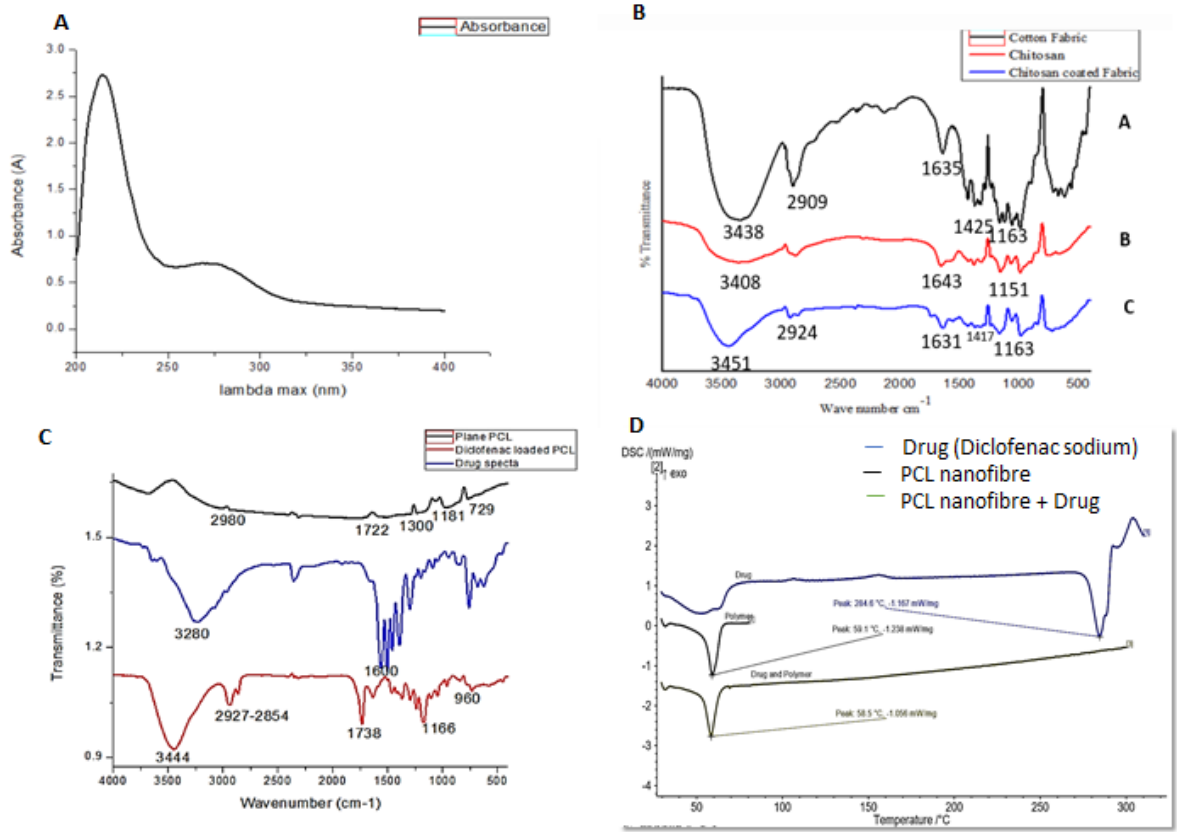

Figure 2 Graph shows the A) UV spectra of drug loaded nanofiber solution B) FTIR spectra of chitosan coated cotton fabric C) FTIR spectra of drug loaded nanofiber D) DSC endothermic peaks of Plane and drug loaded nanofiber

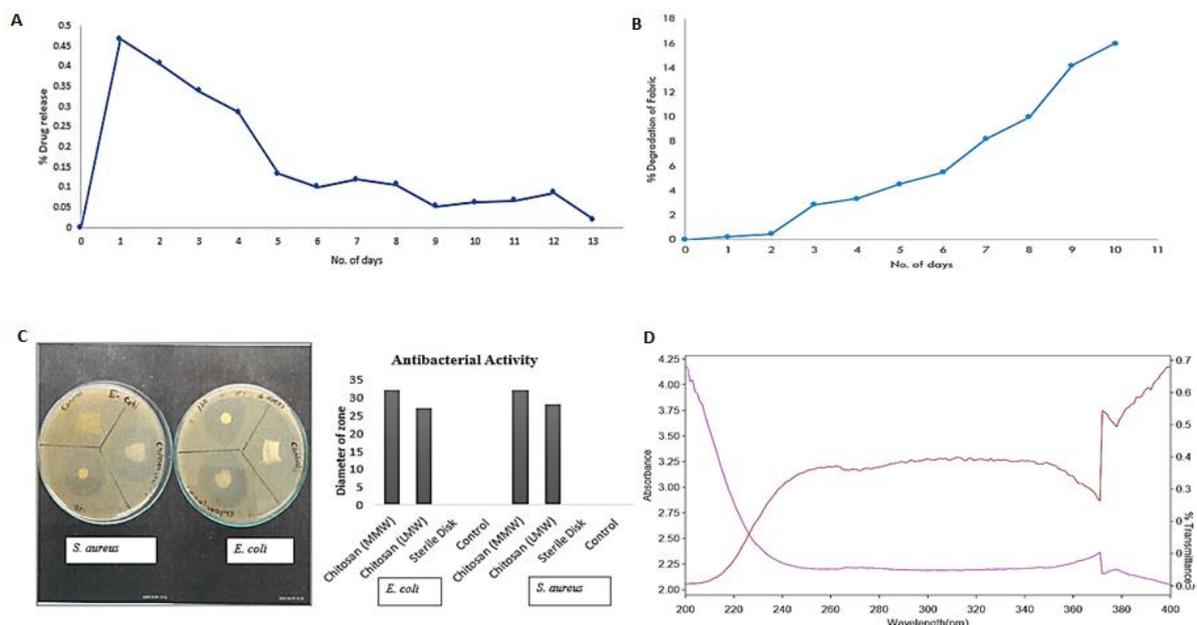

Figure 3 Graph shows that A) Drug release study of nanofiber after every 24 hour B) Degradation study of drug loaded nanofiber film, C) Antibacterial study of chitosan decorated cotton fabric, D) UV blocking activity of ZnO solution

# Supplementary Information

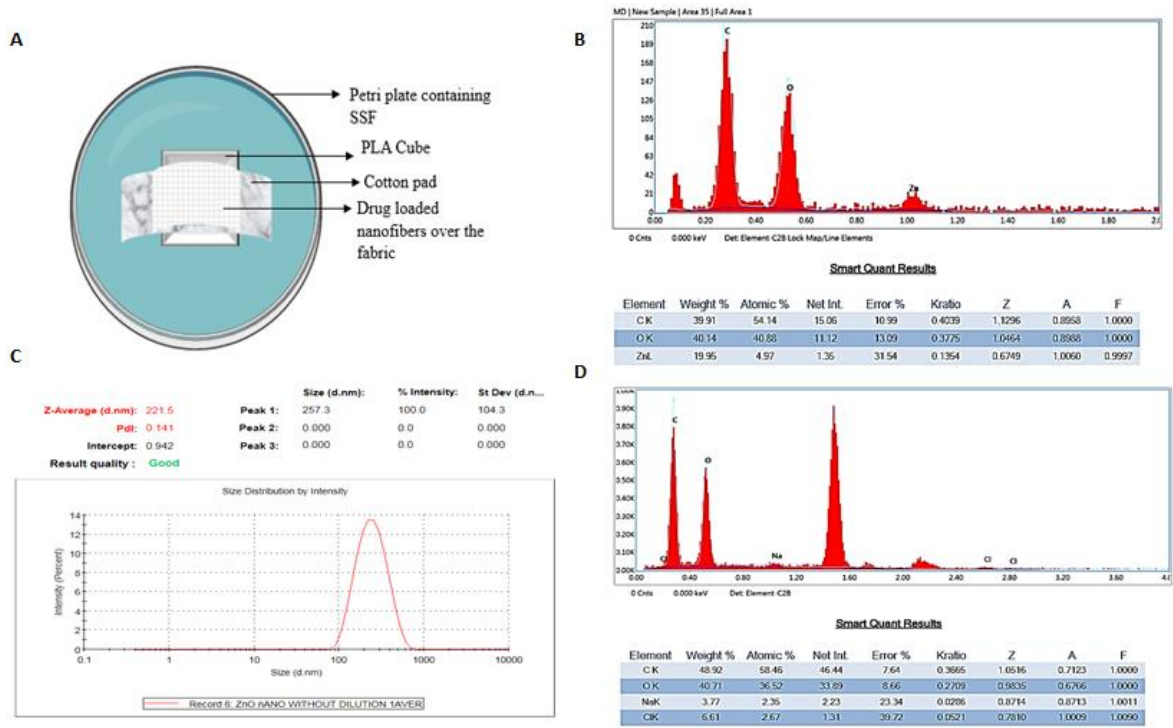

Figure 1: A) Schematic of the experimental setup for the drug release studies for modified CGB B) Elemental analysis of the ZnO coated cotton Gauze bandage C) Particle size distribution of the ZnO nanoparticles D) E Elemental analysis of the diclofenac sodium drug loaded nanofibers